\documentclass[fleqn,10pt]{wlscirep}

\usepackage[utf8]{inputenc}
\usepackage[T1]{fontenc}
\title{Heat, Health, and Habitats: Analyzing the Intersecting Risks of Climate and Demographic Shifts in Austrian Districts}

\author[1]{Hannah Schuster}
\author[2, 1]{Axel Polleres}
\author[2]{Amin Anjomshoaa}
\author[3,4,1,*]{Johannes Wachs}
\affil[1]{Complexity Science Hub, Vienna, AT-1080, Austria}
\affil[2]{Vienna University of Economics and Business, Vienna, AT-1020, Austria}
\affil[3]{Corvinus University of Budapest, Budapest, HU-1093, Hungary}
\affil[4]{HUN-REN Centre for Economics and Regional Studies, Budapest, HU-1097, Hungary}

\affil[*]{johannes.wachs@uni-corvinus.hu}


\begin{abstract}
The impact of hot weather on health outcomes of a population is mediated by a variety of factors, including its age profile and local green infrastructure. The combination of warming due to climate change and demographic aging suggests that heat-related health outcomes will deteriorate in the coming decades. Here, we measure the relationship between weekly all-cause mortality and heat days in Austrian districts using a panel dataset covering $2015-2022$. An additional day reaching $30$ degrees is associated with a $2.4\%$ increase in mortality per $1000$ inhabitants during summer. This association is roughly doubled in districts with a two standard deviation above average share of the population over $65$. Using forecasts of hot days (RCP) and demographics in $2050$, we observe that districts will have elderly populations and hot days $2-5$ standard deviations above the current mean in just $25$ years. This predicts a drastic increase in heat-related mortality. At the same time, district green scores, measured using $10\times 10$ meter resolution satellite images of residential areas, significantly moderate the relationship between heat and mortality. Thus, although local policies likely cannot reverse warming or demographic trends, they can take measures to mediate the health consequences of these growing risks, which are highly heterogeneous across regions, even in Austria.

\end{abstract}
\begin{document}

\flushbottom
\maketitle
%
%
\thispagestyle{empty}

\section*{Introduction}
As global warming due to climate change directly increases the number of heat waves and heat days~\cite{huber_extreme_nodate}, improving heat resilience becomes more crucial than ever. Given the significant increase in extreme heat events observed in Western Europe compared to climate model simulations, there is a deep uncertainty about the health consequences of future summer heat in Europe~\cite{vautard2023heat}. Europe is also rapidly aging, which compounds risks as the elderly are highly vulnerable to hot weather. Many countries in relatively temperate climates are unprepared for high temperatures and require large-scale heat resilience adaptation. The salience of these issues is reflected in the large and growing literature of the effects of heat on human health \cite{iungman_cooling_2023, weilnhammer_extreme_2021} and other aspects of human lives like mobility \cite{massaro_spatially-optimized_2023}. 

However, relatively little work has been done to quantify heat resilience of many comparable regions, municipalities or neighborhoods within a country. For instance, \cite{buzasi_comparative_2022} characterized the heat resilience of districts within Budapest by applying a weighted indicator method, while \cite{zong_mapping_2023} used a directional interaction network to analyze China's heat resilience, revealing indicator directional interactions in the health vulnerability framework, highlighting regional differences within China. Within-country or region analyses of the effect of heat on health outcomes are especially valuable because their units of observation are highly comparable, for instance, because they tend to have similar levels of development and infrastructure, and because it is the level at which most heat resilience interventions are made. At the same time, individual municipalities or regions often lack data or know-how to analyze their vulnerability to warming scenarios, hindering planning efforts.  

In this paper, we study the relationship between heat and all-cause mortality in a panel dataset of all Austrian districts. We show that hot days and heat waves predict significantly higher mortality and that districts with older populations are more vulnerable. On the other hand, above-average concentrations of green vegetation in the residential areas of districts, which we quantify using satellite data, significantly moderates the relationship between health and mortality. This suggests that greening programs are highly effective in improving local heat resilience. We use these results plus district-level forecasts of hot weather and population demographics to generate rankings of heat vulnerability of districts in $2050$ - finding that even the ``coolest'' districts today will be more vulnerable in $2050$ than the ``hottest'' districts today. These estimates, plus our findings on greening, give valuable local knowledge to policymakers about this evolving crisis.

Indeed, multiple factors including climate change, urbanization, and demographic aging are driving an emerging heat and health crisis in Europe. As a result, an increasing number of studies are examining the impact of heat days and heat waves on human health. Additionally, there is growing interest in understanding how urban factors influence the extent of heat stress experienced by individuals, focusing on possible mitigation strategies. Finally, the aging of the population, especially acute in Europe, is likely to amplify the consequences of this crisis~\cite{filiberto2009older}.

The estimated number of heat-related deaths in Europe has risen in the last few years; a study estimated approximately $62,000$ heat-related deaths in Europe between $30$ May and $4$ September $2022$ ~\cite{ballester_heat-related_2023}. It is important to note that the number of heat-related deaths steeply increased with age, and especially women above $80$ years were affected. Furthermore, as heat-related mortality, identified as a critical consequence of climate extremes, rapidly increases and heat-mortality extremes of the past climate are expected to become commonplace, the necessity for adaptation grows~\cite{luethi2023rapid}.

The impact of heat on mortality was found to be mostly immediate, as evidenced by peaks in the risk of death occurring either on the day of exposure or the subsequent day~\cite{Pereira_HeatHealthBurden}. Many studies report a positive association between extreme heat events and cardiovascular or cardiorespiratory mortality ~\cite{weilnhammer_extreme_2021, hajat2005mortality}. Especially a high ratio of elderly people with cardiovascular disease is a common weak point for heat resilience ~\cite{buzasi_comparative_2022}. In Austria, cardiovascular diseases rank as the leading cause of death. Since the probability of cardiovascular diseases increases with age ~\cite{HKE_AUT}, a high ratio of people aged above $65$ years is an additional risk factor for heat resilience.

In addition to its impact on mortality, heat can also adversely affect various other health-related aspects like pregnancies and overall well-being. Several studies have investigated the effect of extreme temperatures on pregnancies. They found that fetal growth is influenced by higher ambient temperatures~\cite{Leung_pregnancy_2022, Pereira_MaternalHealth}, and clinically unobserved pregnancy loss rate increases during extreme heat events~\cite{hajdu_post-conception_2021}. Additionally, extreme heat has also been associated with an increased risk of preterm births ~\cite{cil_extreme_2022}. Overall, heatwaves significantly decrease observed births 9-10 months later \cite{hajdu2024effect}. Furthermore, temperature rises can impact sleep duration, leading to decreased sleep duration ~\cite{Hajdu2023Temperature}, especially among the elderly.

As indicated above, older individuals are particularly vulnerable to the increasing number of heat days, experiencing not only elevated mortality rates but also other adverse effects such as reduced sleep duration induced by heat. This is important as the EU observed an increase in the share of persons aged 65 and above in all member states over the period from $2001$ to $2021$~\cite{EuropeAgeing}. The aging in the EU can be observed in the development of the share of the elderly population, which increased from a population share of $16 \%$ aged over $65$ years in $2001$ to $21 \%$ in $2021$, as well as the increase in the median age from $38$ years in $2001$ to $42$ years in $2011$ and $44$ years in $2021$. Austria is no exception to this trend, with the average age of the Austrian population at $43.2$ years (with Austrian nationals at $45.0$ years and Austrian residents with a different nationality at $35.9$ years)~\cite{demographisches2022}, in $2022$. The population structure of Austria is characterized by a gradual decline in the number and proportion of children and adolescents~\cite{demographisches2022}, which is typical in the EU ~\cite{EuropeAgeing}.

The past two decades demonstrated how challenging the management of adverse weather events is, leading to a heightened vulnerability of populations even in developed European countries. Especially analyzing trends in extreme temperature exposure among European populations reveals a notable surge in heatwave frequency over the last decade, contributing to the increased prevalence of heat-related stress across all cities~\cite{Founda_europeanCities}. Indeed, significant heating effects have been observed even in smaller cities \cite{molnar2020does}. Since more than half of the earth's population currently lives in cities covering less than $3\%$ of the Earth’s land surface ~\cite{massaro_spatially-optimized_2023}, investigating possible interventions becomes more crucial. With the growing world population, this situation will probably get more extreme in the next few years. 

Urban areas are especially at risk due to the urban heat island (UHI) effect and a high population density. Since heatwaves are getting more frequent, stronger, and longer, coupled with the intensifying urbanization further increasing the UHI effect, the thermal risk for urban residents is accelerating~\cite{Founda_Athens,dang2018green}, with measurable effects on mortality. Furthermore, populations residing in areas with high heat exposure predominantly visit locations with similarly high levels of heat, indicating the presence of urban heat traps~\cite{huang_emergence_2023}. Studies find a robust link between greenness and health outcomes in hot weather but tend to focus on large cities \cite{iungman2023cooling}.

As municipal and regional policies can only make marginal contributions to mediate climate change, altering demographic factors like age distribution is not feasible, and with the UHI effect becoming a pressing issue, it is imperative to identify coping strategies that are both easily implementable and fall within municipal or regional budgets. Current research highlights the crucial role of green infrastructure in mitigating urban heat island effects by decreasing exposure to temperature extremes ~\cite{massaro_spatially-optimized_2023} and improving a city’s heat resilience ~\cite{Brile_boosting_2022, WONG2016199}. For instance, ~\cite{iungman_cooling_2023} demonstrated that increasing tree coverage to more than $30\%$ in urban environments not only helps in reducing temperatures but also offers notable health benefits, ultimately fostering the creation of more sustainable and climate-resilient cities.

Our paper studies these three features of the emerging heat and health crisis: a heating climate, an aging population, and local differences in the greenness of built environments. We observe both important cases of heterogeneity and homogeneity across districts. On the one hand, we confirm significant heterogeneities in hot weather, the share of the population over age $65$, and the concentration of greenness in residential areas across Austrian districts observed in $2022$, which correlate significantly with mortality outcomes. Indeed, mortality increases most during heatwaves in districts with older populations, while greener municipalities are less impacted. Turning to forecasts of future warming and aging, we observe a homogeneity: all Austrian districts will get significantly warmer and older.

\section*{Results}
We first estimate the relationship between weekly heat days and death rates for Austrian districts using weekly data in summer months from $2015-2022$. To do so, we use a highly restrictive fixed effects regression, including district, month-of-year, and year-fixed effects. District fixed effects control for time-invariant features of the district, for instance, its location and altitude. Month-of-year fixed effects control for within-summer heat accumulation, and year-fixed effects control for year-specific shocks, for instance, the intensity of infectious diseases like the flu or COVID-19 in a given year. It is especially important to control for the latter because mortality displacement can confound the effect of heat on mortality \cite{hajat2005mortality}. 

\begin{table}
\begingroup
\centering
\begin{tabular}{lcccccc}
   \tabularnewline \midrule \midrule
   Dependent Variable: & \multicolumn{6}{c}{Deaths per 1k Inhabitants (week)}\\
   Model:          & (1)           & (2)           & (3)           & (4)           & (5)           & (6)\\  
   \midrule
   \emph{Variables}\\
   Heat Days       & 0.006$^{***}$ & 0.004$^{***}$ &               &               &               &   \\   
                   & (0.002)       & (0.001)       &               &               &               &   \\   
   Heat Wave       &               &               & 0.017$^{**}$  & 0.011$^{***}$ &               &   \\   
                   &               &               & (0.005)       & (0.003)       &               &   \\   
   Tropical Nights &               &               &               &               & 0.012$^{*}$   & 0.010$^{**}$\\   
                   &               &               &               &               & (0.006)       & (0.003)\\   
   Constant        & 0.169$^{***}$ &               & 0.178$^{***}$ &               & 0.180$^{***}$ &   \\   
                   & (0.005)       &               & (0.005)       &               & (0.005)       &   \\   
   \midrule
   \emph{Fixed-effects}\\
   Month-of-Year           &               & Yes           &               & Yes           &               & Yes\\  
   Year            &               & Yes           &               & Yes           &               & Yes\\  
   District        &               & Yes           &               & Yes           &               & Yes\\  
   \midrule
   Observations    & 6,508         & 6,508         & 6,508         & 6,508         & 6,508         & 6,508\\  
   R$^2$           & 0.011         & 0.175         & 0.008         & 0.173         & 0.011         & 0.175\\  
   Within R$^2$    &               & 0.006         &               & 0.004         &               & 0.007\\  
   \midrule
   \multicolumn{7}{l}{\emph{Clustered (District \& Year) standard-errors in parentheses}}\\
   \multicolumn{7}{l}{\emph{Signif. Codes: ***: 0.01, **: 0.05, *: 0.1}}\\
\end{tabular}
\par\endgroup
\caption{Regression models relating heat variables to mortality outcomes in Austrian districts during June, July, and August, $2015-2022$. }
\label{tab:mainreg}
\end{table}

We report these models in Table~\ref{tab:mainreg} for several heat variables, including the number of heat days (defined as days with $\geq 30$ maximum temperature) in the week, whether there was a heat wave (defined as having three or more heat days), as well as the number of tropical nights (defined as days in which the minimum temperature exceeds $20$ degrees).

We find that controlling for district, month, and year, the marginal effect of an additional heat day increases the number of deaths per $1000$ inhabitants by $0.004$. This is a $2.4\%$ ($0.004$/$0.169$) increase over the general average during summer. The effect is nearly additive: the increase in death rates during heatwaves is nearly three times that of a single heat day ($0.11$ or a $6.5\%$ increase). Finally, tropical nights predict an even greater mortality rate.

\subsubsection*{Heat and Elderly Populations}
There is substantial variation across districts in the share of the population over $65$ (in $2022$: minimum: $15.85\%$, mean: $20.96\%$, maximum: $32.44\%$). In regression models in which we interact heat days with the share of the population over $65$, we observe a significant amplification of the effect of heat on excess deaths in districts with a larger share of elderly inhabitants. We report the results in Table~\ref{tab:elderly_regs}. Note that we include district-level control variables because we can no longer include district-fixed-effects as there is too little variation in the share of the elderly population within districts over the course of our dataset. In particular, models $3$ and $4$ include district-level control variables: average annual income of residents, distance to nearest hospital, and average altitude of residential areas. We report robust (HC) standard errors, noting that our findings are unchanged if we cluster standard errors on year and month. 

\begin{table}
\begingroup
\centering
\begin{tabular}{lcccc}
   \tabularnewline \midrule \midrule
   Dependent Variable: & \multicolumn{4}{c}{Deaths per 1k Inhabitants (week)}\\
   Model:                              & (1)            & (2)            & (3)            & (4)\\  
   \midrule
   \emph{Variables}\\
   Heat Days                           & 0.0056$^{***}$ & -0.0246$^{**}$ & -0.0246$^{**}$ & -0.0232$^{**}$\\   
                                       & (0.0008)       & (0.0115)       & (0.0115)       & (0.0113)\\   
   Share pop. >=65                     & 0.0114$^{***}$ & 0.0079$^{***}$ & 0.0079$^{***}$ & 0.0080$^{***}$\\   
                                       & (0.0007)       & (0.0012)       & (0.0012)       & (0.0011)\\   
   Heat Days $\times$ Share pop. >=65  &                & 0.0015$^{**}$  & 0.0015$^{**}$  & 0.0014$^{**}$\\   
                                       &                & (0.0006)       & (0.0006)       & (0.0006)\\   
   Mean annual gross income (10k Eur)  &                &                & -0.0004        & -0.0168$^{***}$\\   
                                       &                &                & (0.0025)       & (0.0028)\\   
   Distance to nearest Hospital (km)   &                &                &                & -0.0003$^{**}$\\   
                                       &                &                &                & (0.0002)\\   
   Mean Altitude (km)                 &                &                &                & -0.0321$^{***}$\\   
                                       &                &                &                & (0.0038)\\   
   \midrule
   \emph{Fixed-effects}\\
   Year                                & Yes            & Yes            & Yes            & Yes\\  
   Month-of-Year                               & Yes            & Yes            & Yes            & Yes\\  
   \midrule
   Observations                        & 6,508          & 6,508          & 6,508          & 6,508\\  
   R$^2$                               & 0.124          & 0.128          & 0.128          & 0.140\\  
   Within R$^2$                        & 0.113          & 0.117          & 0.117          & 0.129\\  
   \midrule 
   \multicolumn{5}{l}{\emph{Heteroskedasticity-robust standard-errors in parentheses}}\\
   \multicolumn{5}{l}{\emph{Signif. Codes: ***: 0.01, **: 0.05, *: 0.1}}\\
\end{tabular}
\par\endgroup
    \caption{Regression models relating the interaction of heat and share of population over $65$ to mortality outcomes. Heat days are associated with greater mortality when a greater share of a district's population is above $65$. The estimate is robust to including controls for average income and average altitude of the district, as well as year-fixed effects.}
    \label{tab:elderly_regs}
\end{table}

\begin{figure}
    \centering
    \includegraphics[width=0.5\textwidth]{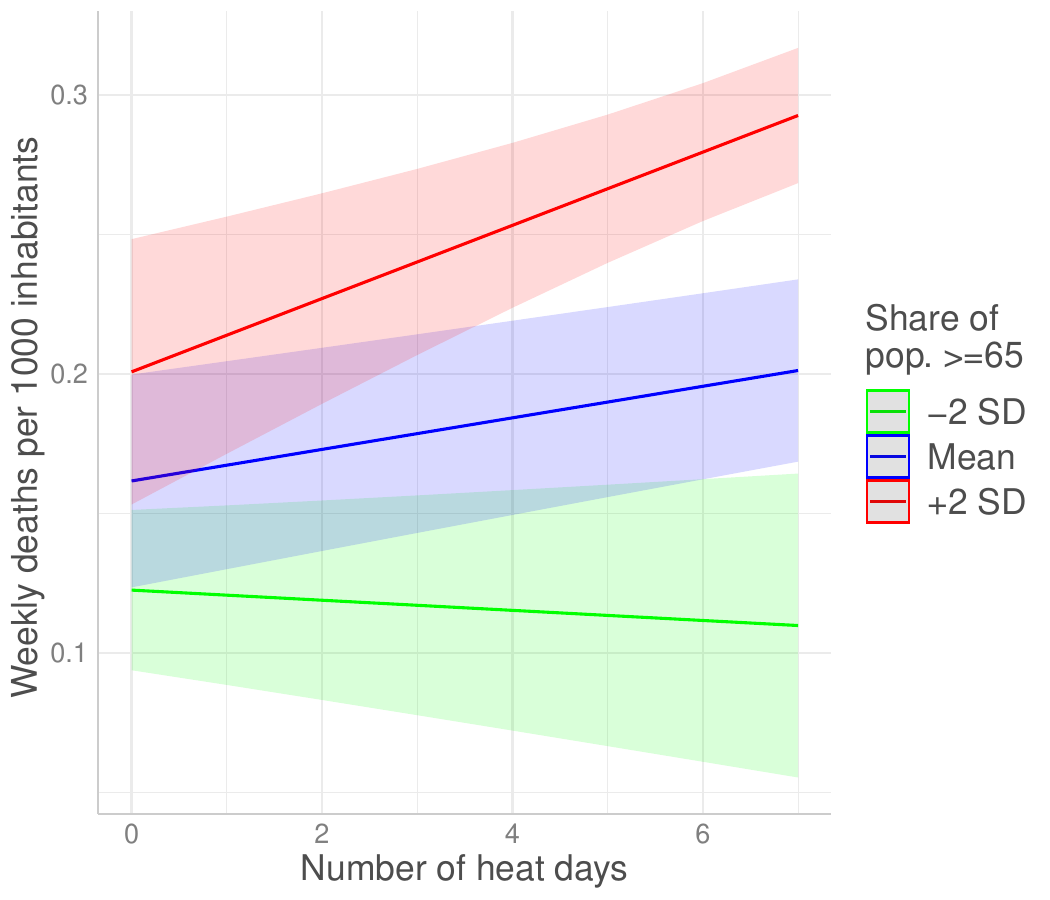}
    \caption{The marginal effects of additional heat days on mortality in Austrian districts, conditioned on the share of the population above $65$. The marginal impact of an additional heat day on mortality is significantly larger in districts with an older population. Estimates are derived from Model $2$ in Table \ref{tab:elderly_regs}, and we report $90\%$ confidence intervals derived from robust standard errors.}
    \label{fig:elderly_intplot}
\end{figure}

We find that the relationship between heat days and mortality is amplified in districts with an older population. This finding is unchanged when adding district-level controls. To better understand the implications of these estimates, we visualize the estimated marginal effects of heat on mortality conditional on the elderly population at the mean and plus or minus two standard deviations in Figure~\ref{fig:elderly_intplot}. The figure suggests that in districts with fewer elderly inhabitants, there is no significant relationship between heat and mortality outcomes. Among the oldest districts, however, heat predicts significantly higher mortality. In the extreme case of a full week of daily maximum temperatures above $30$ degrees, the estimated mortality rate among the oldest districts is nearly twice the average Austria-wide ($0.30$ vs. $0.17$, see Figure ~\ref{fig:elderly_intplot}). 

\subsubsection*{Green Areas}
\begin{figure}
    \centering
    \includegraphics[width=0.75\textwidth]{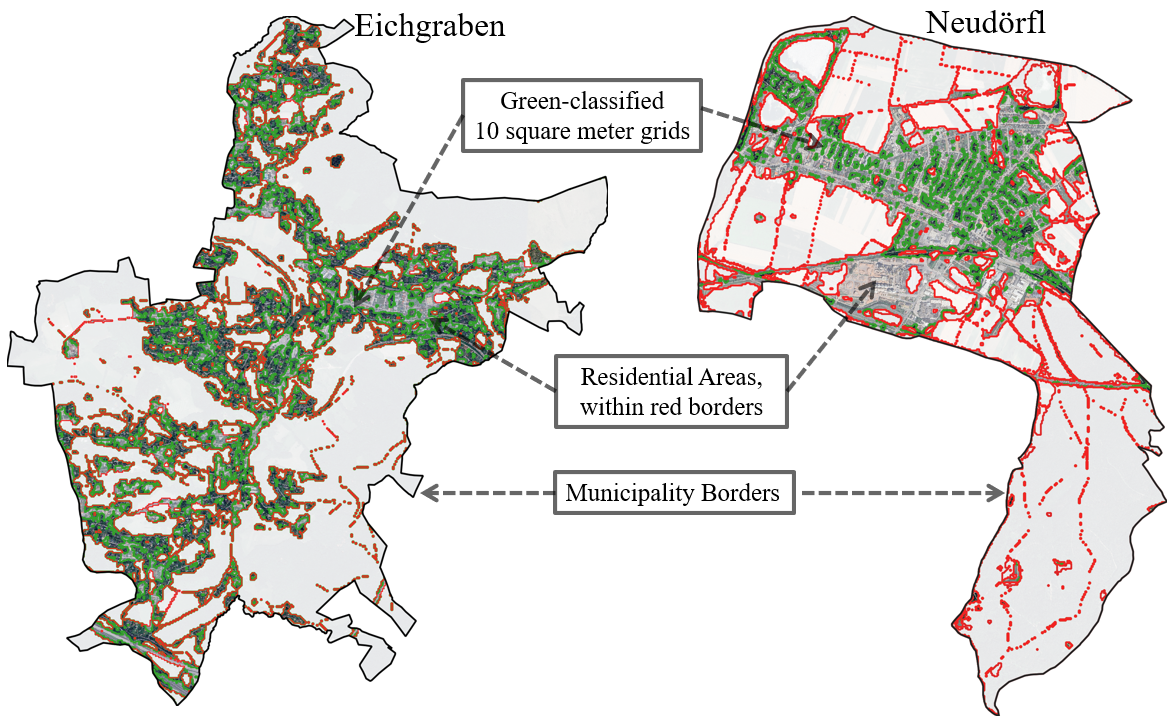}
    \caption{Comparing the greenness of two Austrian municipalities in $2020$, on the left, Eichgraben from Niederösterreich, and on the right, Neudörfl from Burgenland. Both have similar total areas, residential areas, and similar population sizes. However, regarding the greenness of the municipalities using the RGS, Eichgraben has the highest RGS score of all municipalities, while Neudörfl is in the bottom $30\%$ of municipalities ranked by RGS.}
    \label{fig:Greeness}
\end{figure}

Here, we analyze the mediating effect of green vegetation on the relationship between heat and excess mortality. In each district, we consider all residential and commercial areas, defined by a dataset taken from the STATatlas~\cite{Landcover}, ignoring areas outside the built environment. Within these areas, we use the ESA Worldcover V$100$ and V$200$\cite{worldcover2022} satellite data from $2020$ and $2021$, which categorizes areas into $11$ land cover classes using both high-resolution optical Earth observation data from Sentinel-2 and SAR (Synthetic Aperture Radar) data from Sentinel-1 at a resolution of $10$ square meters. We derive a greenness score from the data by considering the relative share of green-classified squares to the total area and denoted it as Residential Green Share (RGS). We show two examples of very different RGS in Figure \ref{fig:Greeness}, emphasizing that we only consider areas in the built environment. We chose these two municipalities because they are similar in many aspects:
\begin{itemize}
    \item They have similar areas (Neudörfl: $9.02$ km$^2$, Eichgraben:$8.88$ km$^2$),
    \item of which a similar area is residential (Neudörfl: $3.12$ km$^2$, Eichgraben:$3.16$ km$^2$),
    \item and their populations are similar (Neudörfl: 4 641, Eichgraben:4 652).
\end{itemize}
However, comparing the RGS value we use to measure the greenness of the municipality, reveals that the two have significantly different levels of green infrastructure in their residential areas (Neudörfl: $23.36 \%$, Eichgraben:$72.42\%$). Specifically, Eichgraben has the highest RGS score of all Austrian municipalities, while Neudörfl is in the bottom $30\%$.

Returning to the regression model framework, we again model weekly deaths per $1000$ inhabitants as a function of heat, using the strong heatwave indicator as we only have fine-grained satellite data for two years. We standardize the greenness score (mean $0$, standard deviation $1$). We find that the relationship between heat waves and increased mortality is mediated in districts with one standard deviation above average greenness score. This result holds even when controlling for the district's share of the population over $65$, average income, and mean altitude.

\begin{table}[]
\begingroup
\centering
\begin{tabular}{lcccc}
   \tabularnewline \midrule \midrule
   Dependent Variable: & \multicolumn{4}{c}{Deaths per 1k Inhabitants (week)}\\
   Model:                              & (1)           & (2)           & (3)            & (4)\\  
   \midrule
   \emph{Variables}\\
   Heat Wave                           & 0.013$^{***}$ & 0.011$^{***}$ & 0.012$^{***}$  & 0.012$^{***}$\\   
                                       & (0.005)       & (0.004)       & (0.004)        & (0.004)\\   
   Greenness Score                     & -0.004        & 0.002         & 0.001          & 0.003\\   
                                       & (0.003)       & (0.002)       & (0.002)        & (0.002)\\   
   Heat Wave $\times$ Greenness Score  &               & -0.018$^{**}$ & -0.019$^{**}$  & -0.015$^{*}$\\   
                                       &               & (0.009)       & (0.009)        & (0.009)\\   
   Mean annual gross income (10k Eur)  &               &               & -0.018$^{***}$ & -0.013$^{***}$\\   
                                       &               &               & (0.005)        & (0.005)\\   
   Share pop. >=65                     &               &               &                & 0.011$^{***}$\\   
                                       &               &               &                & (0.001)\\   
   Mean altitude (km)                  &               &               &                & -0.022$^{***}$\\   
                                       &               &               &                & (0.005)\\   
   \midrule
   \emph{Fixed-effects}\\
   Year                                & Yes           & Yes           & Yes            & Yes\\  
   Month                               & Yes           & Yes           & Yes            & Yes\\  
   \midrule
   Observations                        & 2,261         & 2,261         & 2,261          & 2,261\\  
   R$^2$                               & 0.024         & 0.033         & 0.038          & 0.136\\  
   Within R$^2$                        & 0.007         & 0.017         & 0.022          & 0.121\\  
   \midrule 
   \multicolumn{5}{l}{\emph{Heteroskedasticity-robust standard-errors in parentheses}}\\
   \multicolumn{5}{l}{\emph{Signif. Codes: ***: 0.01, **: 0.05, *: 0.1}}\\
\end{tabular}
\par\endgroup
    \caption{Regression models investigating how district greenness within residential areas mediates the relationship between heat and mortality. Note greenness scores are only available in $2020$ and $2021$.}
    \label{tab:green_regs}
\end{table}

\subsection*{Forecasts}
\begin{figure}
    \centering
    \includegraphics[width=0.75\textwidth]{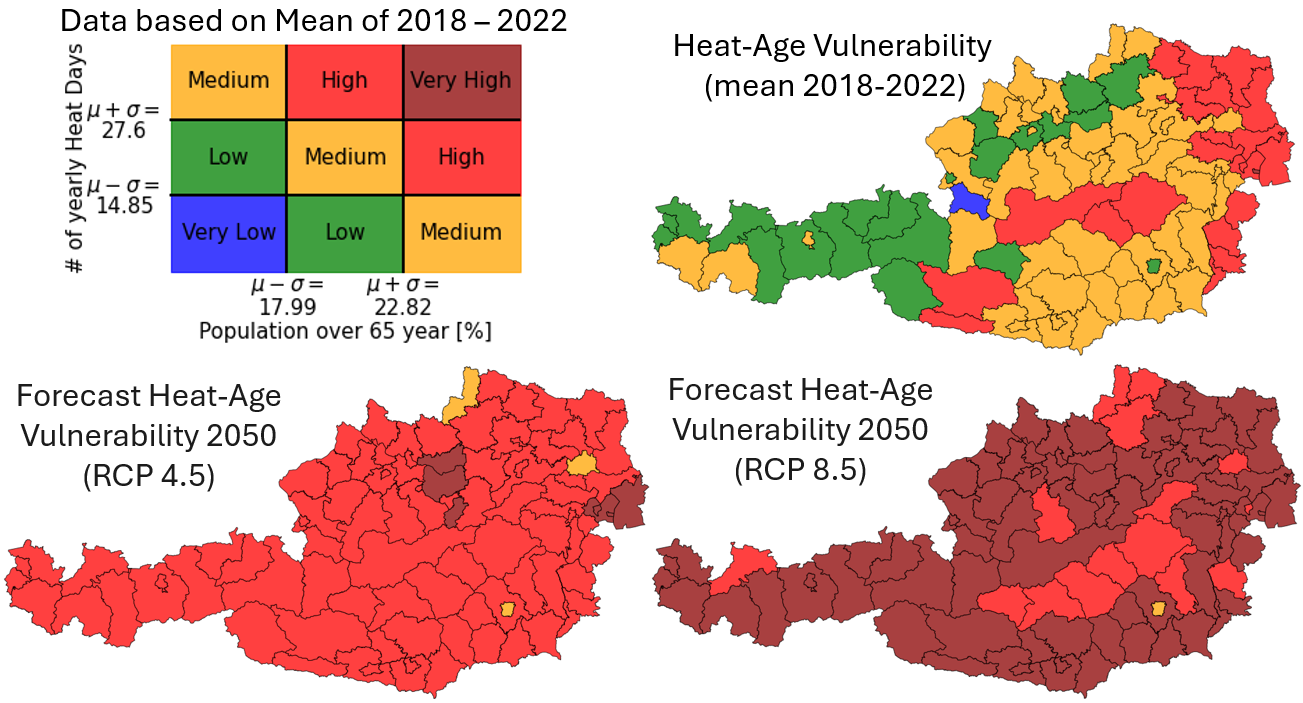}
    \caption{In this figure, we compare the change of the heat-age vulnerability of Austrian districts using mean data over the last $5$ years for the current situation and forecast data for $2050$ under the RCP $4.5$ and RCP $8.5$ scenario.}
    \label{fig:MapCompare}
\end{figure}

So far, we have shown that hot weather predicts higher mortality rates, that elderly people are more vulnerable, and that the density of green vegetation in residential areas may mitigate this effect. We now use forecasts of hot weather and demographics to estimate heat-related health risks in Austria in $2050$. Austria, like many developed countries, is aging rapidly. District-level forecasts also suggest significant warming will increase heat days and heat waves. In Figure \ref{fig:MapCompare}, we visualize the recent and forecast number of weekly heat days and the current and forecast share of the population of $65$ years for all Austrian districts. 

\begin{figure}
    \centering
    \includegraphics[width=0.75\textwidth]{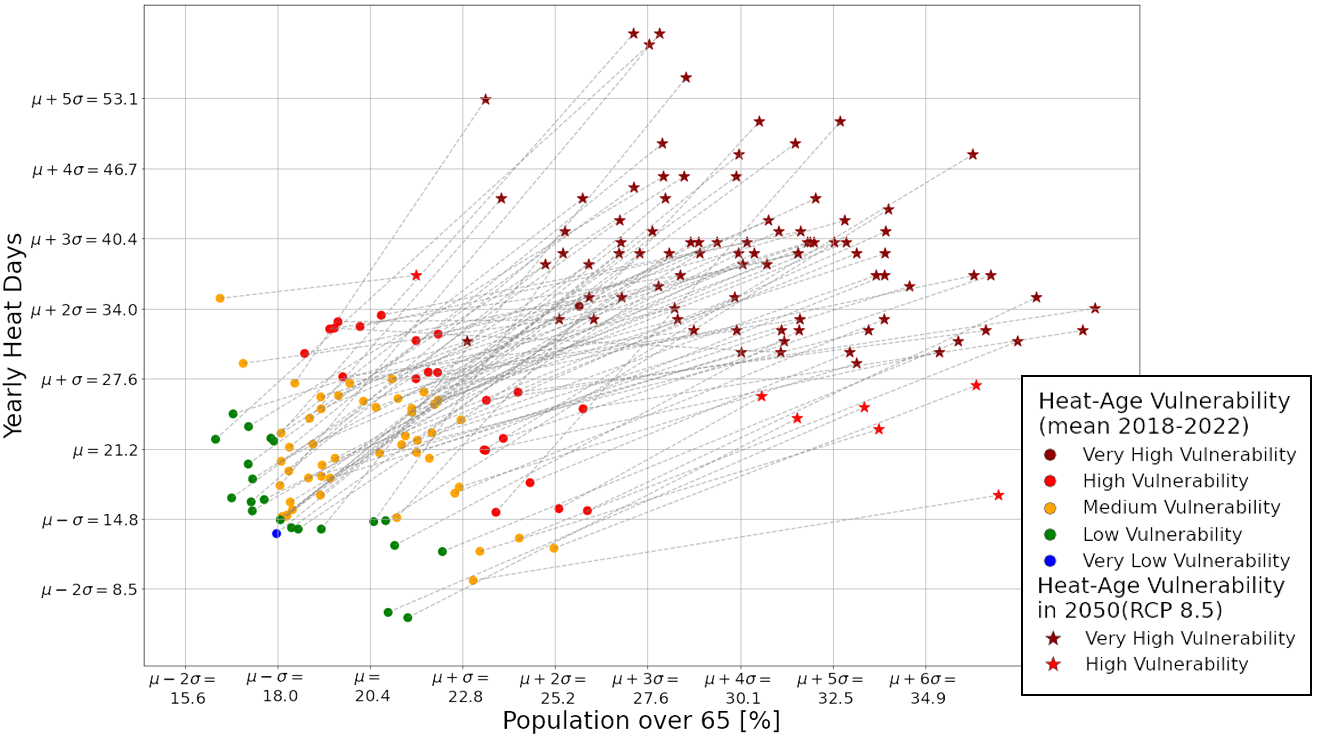}
    \caption{In this plot, we made the shift in the number of yearly heat days and the change in population share over $65$ years visible by comparing the mean heat-age vulnerability over the last $5$ years from $2018$ to $2022$ (dots) vs. the projected heat-age vulnerability in $2050$ under the RCP $8.5$ scenario (stars).}
    \label{fig:MovingCompare}
\end{figure}

We observe that even the youngest and more temperate districts in Austria in $2050$ will resemble a top $20\%$ district in $2022$ in terms of heat-age vulnerability. We visualize the shift of each district in the heat-agedness space from $2022$ to $2050$ in Figure \ref{fig:MovingCompare}. The axis ticks mark the average and standard deviations of both distributions in the $2018-2022$ data.



\section*{Discussion}
In this paper, we demonstrated a significant variation in population heat vulnerability within Austria. We showed that heat is correlated with a significant increase in mortality, especially in districts with a larger share of the elderly population. Forecasts suggest that by $2050$, nearly all Austrian districts will be in the top $20\%$ of heat and age risk of 2022. There is one exception: Murau, a district located in the Styrian Alps, which is forecast to be in the $65$th percentile by $2022$ standards in $2050$. The strong correlations we have found between heat and mortality, amplified in older populations, underscore the health risks of this potential future. 

However, we also observe that the greenness of the built environment significantly mitigates the relationship between heat and mortality. While local governments and initiatives can only make marginal contributions to the fight against global warming, they have a significantly greater ability to increase cooling vegetation in their built environment. Such investments have the advantage of being cumulatively and approximately linearly effective: cities can be made green block by block without a large scale up-front cost. Public investments can influence private choices, such as the proliferation of air conditioning \cite{davis2015contribution}. The growing health risks of a warming planet and an aging population suggest that cost-benefit calculations should be used to make decisions about public investments. 

Such calculations require more fine-grained and accurate forecasts of both weather and demographic trends. For example, our results are dependent on forecasts of future extreme heat events, which have recently been criticized as underestimates by some~\cite{vautard2023heat}, and overestimates by others~\cite{brunner2024pitfalls}. Furthermore, it is generally difficult to estimate the effectiveness of green infrastructure projects \cite{heidari2022cost}.

In general, a precise estimate of the social costs of heat on health is beyond the scope of this paper as we only consider the extreme outcome of mortality. In reality, increased use of the healthcare system and decreases in productivity during hot weather are likely to be very costly, suggesting that even extensive efforts to cool built environments will pay off in the long run.

\section*{Methods}
In this section, we will introduce the study area and discuss the various datasets utilized, including how they were obtained and aggregated, before delving into an explanation of the regression model employed.
\subsection*{Study Area} 
Austria is a small, mountainous, landlocked country located in southern central Europe, covering part of the eastern Alps and the Danube region\cite{austria_nc7}. It is a highly developed country with an HDI value of $0.926$ and an HDI rank of $22$ of $193$ countries~\cite{undp_hdi} in $2022$. Access to healthcare in Austria is not considered an issue as the level of unmet medical needs is low, and it has the second-highest density of doctors in the EU in $2021$~\cite{healthcareAUT}, but also the lowest proportion of general practitioners in the EU. However, Austrian healthcare is one of the most expensive healthcare systems in the EU, with the third highest spending per capita in $2021$. But still, the health system remains structurally and financially fragmented, with an imbalance between regions, medical specialties, and an aging physician workforce. This poses a problem as the demand for healthcare is increasing due to population aging and an increasing life expectancy at $81.1$ years in $2022$, which is $0.4$ years higher than the EU average. In $2022$, the average age of the Austrian population was at $43.2$ years (with Austrian nationals at $45.0$ years and Austrian residents with a different nationality at $35.9$ years)~\cite{demographisches2022}. Furthermore, Austria has a low fertility rate, which was at $1.48$ children per woman in $2021$. Consequently, for several decades, the population structure of Austria has been characterized by a gradual decline in the number and proportion of children and adolescents, accompanied by an increase in the number and proportion of elderly individuals, resulting in population aging~\cite{demographisches2022}.

The climate in Austria is notably heterogeneous due to its unique position within the central European transitional climatic zone, heavily influenced by its rugged topography, particularly the Alps~\cite{austria_nc7}. The country can be divided into three main climatic regions: 
\begin{itemize}
\item The continental Pannonian climate in the eastern part is characterized by a mean temperature during July of usually above $19$°C and an annual precipitation level often less than $800$ mm.
\item The Alpine Climate in the central Alpine region is known for its high precipitation levels, short summers, and long winters.
\item The transitional central European climatic zone in the rest of Austria features a wet and temperate climate with mean July temperatures of $14–19$°C and annual precipitation levels ranging from $700–2000$ mm, depending on location, exposure, and altitude.
\end{itemize}



\subsection*{Data}
In this paper, we gathered data from various sources to comprehensively analyze the heat resilience of municipalities. It is important to note that these data sets originate from diverse sources, resulting in variations in spatial granularity for various reasons. This subsection will describe the data sets themselves, including the aggregation method used for our regression models. Given the diverse sources of our datasets, we developed a Knowledge Graph—a specialized form of database—to integrate the portion of our datasets accessible through open-source platforms~\cite{anjomshoaa2023data}. This approach prioritizes transparency and enhances user accessibility.

\paragraph{Geography} Due to the spatial nature of our analysis, we start by defining the different spatial levels. The datasets we use in our analysis encompass various spatial levels, ranging from the granular square kilometer level to the broader Austrian administrative units of municipalities and their higher-level districts taken from open.data by Statistics Austria~\cite{muniborders, distborders}. We began our data collection process by incorporating the borders of municipalities, which is crucial for defining the number of heat days and identifying areas of interest within them. Subsequently, we introduced the districts' borders one level above the municipality level. This decision was driven by the fact that the number of weekly deaths is only accessible at the district level due to data protection regulations.

We initiate the data aggregation process by standardizing all datasets to a uniform level, specifically the municipality level. The three main methods we employed are maximum, mean, and calculating the proportion or percentage. Utilizing percentages allows for comparability among different municipalities, as the data is expressed in a standardized way. In our regression model, the dependent variable is on the district level, while our independent variables are on the municipality level. We use a population-weighted mean to aggregate the independent variable from the municipality level to the district level. This decision was made because the death rate, which is the independent variable in our regression model, heavily depends on the population of the different municipalities located within the district. Consequently, we wanted to ensure that the traits of municipalities with a larger population have a higher impact on the death rate. This translates to the following formula:
\begin{equation}
    X_{D} = \sum_{M \in D}  \frac{X_M * P_M}{P_D},
\end{equation}
here the aggregate value $X_{D}$ on district level $D$ is calculated by summing the product of $X_M$ and its population  $P_M$ for each municipality $M$ within district $D$, weighted by the  total population $P_D$ of district $D$. This population-weighted aggregation method is consistently applied in this paper unless specified otherwise.


\paragraph{Heat and Weather} The second dataset for this analysis is the number of weekly heat days per municipality. A heat day is defined as a day with a temperature exceeding $30$ °C. We aggregated this data from the Spartacus Dataset ~\cite{SPARTACUS}, which is on km$^2$, to the municipality level by calculating the maximum temperature. Even though the data set has temperature data since $1961$, for this analysis, we will only focus on the time frame from $2015$ to $2022$. The air temperature in this dataset was accumulated by the Austrian weather agency Geosphere and is measured at their stations at $2$ meters above the ground.

\paragraph{Demographics} One significant risk factor when speaking about heat resilience is the population. In the Literature especially, people above the age of $65$ are counted as at-risk. We decided to use a percentage of the whole population to characterize how many people are above the age of $65$; the data is available on the STATatlas provided by Statistics Austria ~\cite{StatAtlas} [Section Population - Subsection Population by age - Variable: Age 65 years and older in \%] and covers the time frame from $2002$ to $2022$. Furthermore, we were able to obtain a dataset with weekly deaths per district directly from Statistics Austria. Even though a district is a level above the municipality, we were able to use the provided data to validate the impact of heat days on the population. 

\paragraph{Greenness} There are many ways to improve a municipality's heat resilience. However, given the constraints of a limited budget, assessing the effectiveness of these measures takes on added significance. One possible measure is to improve the availability of green spaces in the residential area, which is proven to decrease the temperature during the summer. For this, we needed to define a measurement that can be used to quantify the greenness of every municipality in a manner that allows for a comprehensive assessment of its impact on the heat during summer.

In Austria, many municipalities consist of residential areas surrounded by natural areas, and in many cases, they are covered in woods. However, these areas can improve the liveability of the population only to a certain extent because their positive effect is limited by distance. Consequently, we refer to greenness as the amount of plant life, like trees within the residential areas providing shade and cooling the environment. For this purpose, we needed a data set that would enable merging with administrative borders and greenness filtering.

For the residential Area in Austria, we used a data set provided by Statistics Austria that contains land use information~\cite{Landcover} and extracted polygons and multipolygons that encompass the residential areas using QGIS. In the next step, the data was filtered using the Python library geopandas to get a dataset containing each municipality's residential and commercial area. The green area is filtered from the ESA Worldcover V100 and V200, \cite{worldcover2022}, providing a dataset on land use data based on satellite images. The Austrian border was used to extract the area of interest using QGIS, and afterward, the different polygons and multipolygons were assigned to the different municipalities. Calculating the intersection of these two data sets provided us with a dataset containing the greenness of every municipality in the form of polygons for each municipality. We then calculated the proportion of green area for every municipality and denoted it as Residential Green Share (RGS).

\paragraph{District Control Factors}
We use different district control factors in our regression model, obtained from different sources and aggregated to the district level if necessary. We use the average gross earnings of employees with full-year earnings from Statistics Austria ~\cite{StatAtlas} [Section: Public Finances and Taxes - Subsection: Wage tax statistics - Variable: Employees]. We include this in our data set using a population-weighted mean. The mean altitude was at first calculated for each municipality using the Spartacus data set from Geosphere ~\cite{SPARTACUS} and aggregated to the district level using a population-weighted mean. Furthermore, we calculated the distance to the closest hospital in \cite{schuster2024stress}, which we also include in this analysis. The initial data set is on the municipality level and is aggregated to the district level using a population-weighted mean.



\paragraph{Heat and Demographic Forecast Data}
In this section, we present the forecast data utilized in our analysis. We provide an overview of the sources and methodologies employed to capture future trends and patterns, enabling a comprehensive understanding of the data under examination.

The demographic forecast data set we use predicts the population changes for Austria at the district level, as provided by Statistics Austria ~\cite{forecastdemographic}. This is the smallest spatial unit Statistics Austria uses for its demographic forecasts. We focus here again on the population share over $65$ years. 

The weather data is from a Geosphere project called OEKS15~\cite{oeks15}, which uses different modeling approaches to predict yearly heat days on a square km grid and was provided by them directly. We decided to use 'ICHEC-EC-EARTH\_r12i1p1\_\newline SMHI-RCA4', an EC-EARTH global climate model downscaled with the SMHI-RCA4 regional climate model, which is the fourth version of the Rossby Centre Regional Atmospheric Climate Model [RCA4] from the Swedish Meteorological and Hydrological Institute [SMHI]. The dataset is on a km$^2$ level and consists of yearly heat days for different climate scenarios. Since the demographic forecast is available only at the district level, we directly calculated the number of yearly heat days for the two different scenarios at the district level.

For the prediction of heat days in the near future, we used two different scenarios: the RCP$4.5$, which corresponds to the Klimaschutz-Szenario (climate change mitigation scenario), and the RCP$8.5$, which models business-as-usual. In the near future, both scenarios predict similar outcomes when comparing the results on average, with an increase of $11$ summer days and $4.3$ heat days. The signal of change is especially significant in lower-lying areas. By the end of the $21$st century, the difference between the moderate RCP$4.5$ scenario and the more extreme RCP$8.5$ scenario becomes more apparent. The RCP$4.5$ scenario projects an increase of $18$ summer days (with a range of $13.1$ to $29.8$ days) and $7.0$ heat days (with a range of $4.6$ to $13.1$ days), while the RCP$8.5$ scenario shows an average increase of $35$ summer days (with a range of $25.4$ to $55.6$ days) and $17.4$ heat days (with a range of $11.2$ to $32.4$ days).

\subsection*{Models}
We fit multiple linear regression models using Ordinary Least Squares (OLS) to estimate the relationship between hot weather and mortality at the district-week level. In our first models, we use fixed-effect heavy specifications to control for time-invariant district factors, as well as year and month-of-year invariant factors. More specifically, given district $d$ and week $t$, we estimate:

$$
\text{Death\_rate\_per\_1000}_{d,t} = \beta_0 + \beta_1 \text{Heat\_Days}_{d,t} + \mu_{\text{Year}} + \eta_{\text{Month-of-Year}} +\gamma_{d} + \varepsilon
$$

where $\beta_1$ estimates the marginal effect of an additional heat day on mortality per $1,000$ inhabitants that week. Other specifications substitute tropical nights or heat waves (a binary variable defined as $1$ if at least $3$ hot days are observed in the week) for heat days. We cluster standard errors on district and year.

In subsequent models, we are interested in estimating the effect of the interaction of district-specific variables with heat days (or other heat observations). Specifically, we are interested in either the share of the population over $65$, or the green-score of the district, both observed at an annual level. In both cases, we do not have enough variation over the course of our data to include district fixed effect. The high-resolution satellite data used to measure green scores is only available for $2020$ and $2021$, and the share of the elderly population changes in a relatively steady and homogeneous way in Austrian districts $2015-2022$. Thus, we estimate (in one case):

$$
\text{Death\_rate\_per\_1000}_{d,t} = \beta_0 + \beta_1 \text{Heat\_Days}_{d,t} + \beta_2 \text{Share of Population over 65} + \beta_3\text{I(Heat,Elderly)} + \textbf{X} + \mu_{\text{Year}} + \eta_{\text{Month-of-Year}}+ \varepsilon
$$

where I(Heat,Elderly) is the interaction of the heat days and share of the population over $65$, and \textbf{X} is a matrix of district-level controls (income per capita, altitude, etc.). The models fit to study the greenness score are similar. We report robust standard errors for all of these models (i.e. excluding district-fixed effects).

\section*{Acknowledgements}
The authors acknowledge support from the Austrian Federal Ministry for Climate Action, Environment, Energy, Mobility and Technology (BMK) via the ICT of the Future Program - FFG No 887554. JW acknowledges support from the Center for Collective Learning (101086712-LearnData-HORIZON-WIDERA-2022-TALENTS-01 financed by EUROPEAN RESEARCH EXECUTIVE AGENCY (REA)). We thank Elias Wellems and Matthias Humer for their assistance with satellite data. We thank David Idl for his exploring this topic in his bachelor thesis. We thank Geosphere and Statistic Austria for their assistance in providing the datasets.

\bibliography{sample}

\end{document}